\documentclass[reprint, amsmath,amssymb, aps, prb,twocolumn,superscriptaddress]{revtex4-1}
\usepackage{graphicx}% Include figure files
\usepackage{dcolumn}% Align table columns on decimal point
\usepackage{bm}% bold math

\usepackage{color}

\begin{document}

\title{Magnetic properties of nanoparticles compacts  with controlled broadening of the particle size distribution}
%\title{Magnetic properties of compacts of maghemite nanoparticles with controlled broadening of the particle size distribution}
%\title{Effect of the controlled broadening of the nanoparticle size distribution onto the dipolar superspin glass phase transition\\ alt.\\ Dipolar superspin glass phase transition in compacts of nanoparticles with controlled broadening of the size distribution}%
%\title{Controlled broadening of the particle size distribution in a dense ensemble of maghemite nanoparticles and superspin glass properties}%
%\shorttitle{Superspin glasses with varying particle size distribution} %Insert here a short version of the title if it exceeds 70 characters

%\author{Mikael Svante Andersson\inst{1} \thanks{E-mail: \email{Mikael.Andersson@angstrom.uu.se}} \and Jose Angel De Toro\inst{2} \and Su Seong Lee\inst{3} \and Roland Mathieu\inst{1} \and Per Nordblad\inst{1}}
%\shortauthor{Mikael Svante Andersson \etal}

%\institute{                    
 % \inst{1} Department of Engineering Sciences, Uppsala University, Box 534, SE-751 21 Uppsala, Sweden\\
  %\inst{2} Instituto Regional de Investigaci\'on Cient\'ifica Aplicada (IRICA) and Departamento de F\'isica Aplicada, Universidad de Castilla-La Mancha 13071 Ciudad Real, Spain\\
	%\inst{3} Institute of Bioengineering and Nanotechnology, 31 Biopolis Way, The Nanos, Singapore 138669, Singapore
	
%	}
%\pacs{5.50.Tt}{Fine-particle systems; nanocrystalline materials}
%\pacs{75.50.Lk}{Spin glasses and other random magnets}
%\pacs{75.75.-c}{Magnetic properties of nanostructures}

\author{M. S. Andersson}
\email{Mikael.Andersson@angstrom.uu.se}
\affiliation{Department of Engineering Sciences, Uppsala University, Box 534, SE-751 21 Uppsala, Sweden}

\author{R. Mathieu}
\affiliation{Department of Engineering Sciences, Uppsala University, Box 534, SE-751 21 Uppsala, Sweden}

\author{P. S. Normile}
\affiliation{Instituto Regional de Investigaci\'on Cient\'ifica Aplicada (IRICA) and Departamento de F\'isica Aplicada, Universidad de Castilla-La Mancha, 13071 Ciudad Real, Spain}

\author{S. S. Lee}
\affiliation{Institute of Bioengineering and Nanotechnology, 31 Biopolis Way, The Nanos, Singapore 138669, Singapore}

\author{G. Singh}
\affiliation{Department of Materials Science and Engineering, Norwegian University of Science and Technology (NTNU), 7491Trondheim, Norway}

\author{P. Nordblad}
\affiliation{Department of Engineering Sciences, Uppsala University, Box 534, SE-751 21 Uppsala, Sweden}

\author{J. A. De Toro}
\affiliation{Instituto Regional de Investigaci\'on Cient\'ifica Aplicada (IRICA) and Departamento de F\'isica Aplicada, Universidad de Castilla-La Mancha, 13071 Ciudad Real, Spain}

\begin{abstract}
Binary random compacts with different proportions of small (volume V) and large (volume 2V) bare maghemite nanoparticles (NPs) are used to investigate the effect of controllably broadening the particle size distribution on the magnetic properties of magnetic NP assemblies with strong dipolar interaction. A series of eight random mixtures of highly uniform 9.0 and 11.5 nm diameter maghemite particles prepared by thermal decomposition are studied. In spite of severely broadened size distributions in the mixed samples, well defined superspin glass transition temperatures are observed across the series, their values increasing linearly with the weight fraction of large particles.
\end{abstract}

\maketitle

\section{Introduction}
Magnetic nanoparticle (NP) assemblies are the basis of an ever-increasing catalogue of applications.\cite{Deng2015,Jordan2007,Salata2004} Intuitively, it is expected that the nanoparticle size distribution will affect the properties of a dense magnetic particle ensemble, e.g., by allowing or forbidding the development of critical-slowing-down and a superspin glass phase transition. However, there is a scarcity of experiments specifically designed to verify and quantify such effects. On the other hand, a vast literature on strongly interacting (dense) particle systems suggests that increasing the width of a NP size distribution will eventually  destroy the collective spin-glass-like transition \cite{Jonsson1995PRL,Hansen2002,Peddis2014Book} or yield "percolation" transitions where only a fraction of the NP superspins freeze cooperatively.\cite{Parker2008}

Current synthesis technology has allowed the preparation of highly uniform NPs, enabling the formation of self assembled NP analogues of atomic compounds.\cite{Disch2011,Chen2010,Yang2015} Here, random binary compacts are prepared with different proportions of small (9.0 nm in diameter) and large (11.5 nm) bare maghemite NPs as a method to both fine tune the average particle moment and control the particle size distribution in order to study the influence of both factors on the collective magnetism of the ensembles. The difference in particle diameter corresponds to a factor 2 difference in volume. The end members of this binary series (made solely of small or large particles) have been shown to exhibit sharp superspin glass (SSG) transitions (due to strong dipolar interactions) at temperatures much larger than the individual particle blocking temperatures.\cite{Andersson2015Nanotechnology, Andersson2016MRX}

\begin{figure*}[t!]
	\centering
		\includegraphics[width=0.95\textwidth]{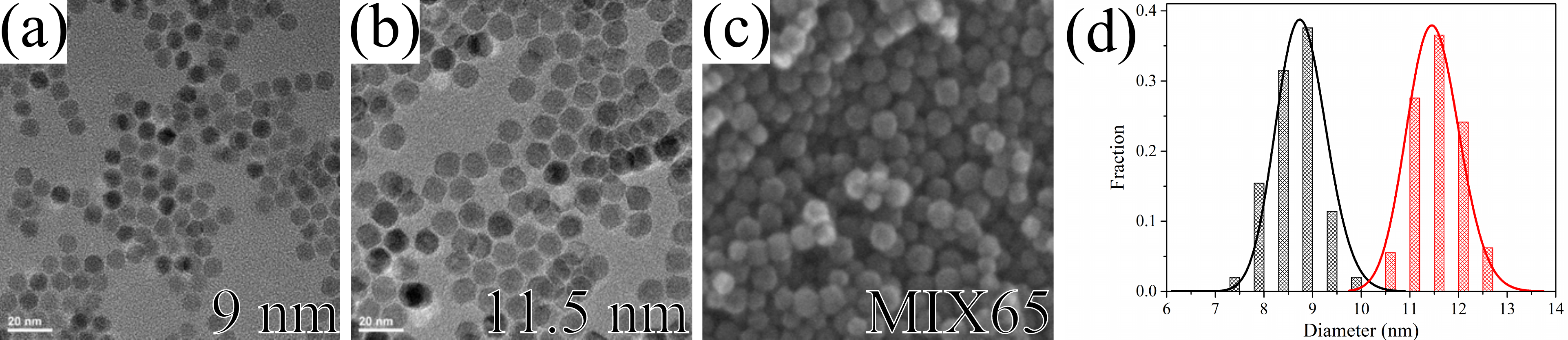}
	\caption{TEM images of 9 nm (a) and 11.5 nm (b) particles used to make all of the MIXx compacts. (c) A representative SEM image of a MIXx sample (x=65). (d) Size distributions of the 9 and 11.5 nm particles.} 
	\label{fig:TEM_SEM}
\end{figure*}

\begin{figure*}[t!]
	\centering
		\includegraphics[width=0.950\textwidth]{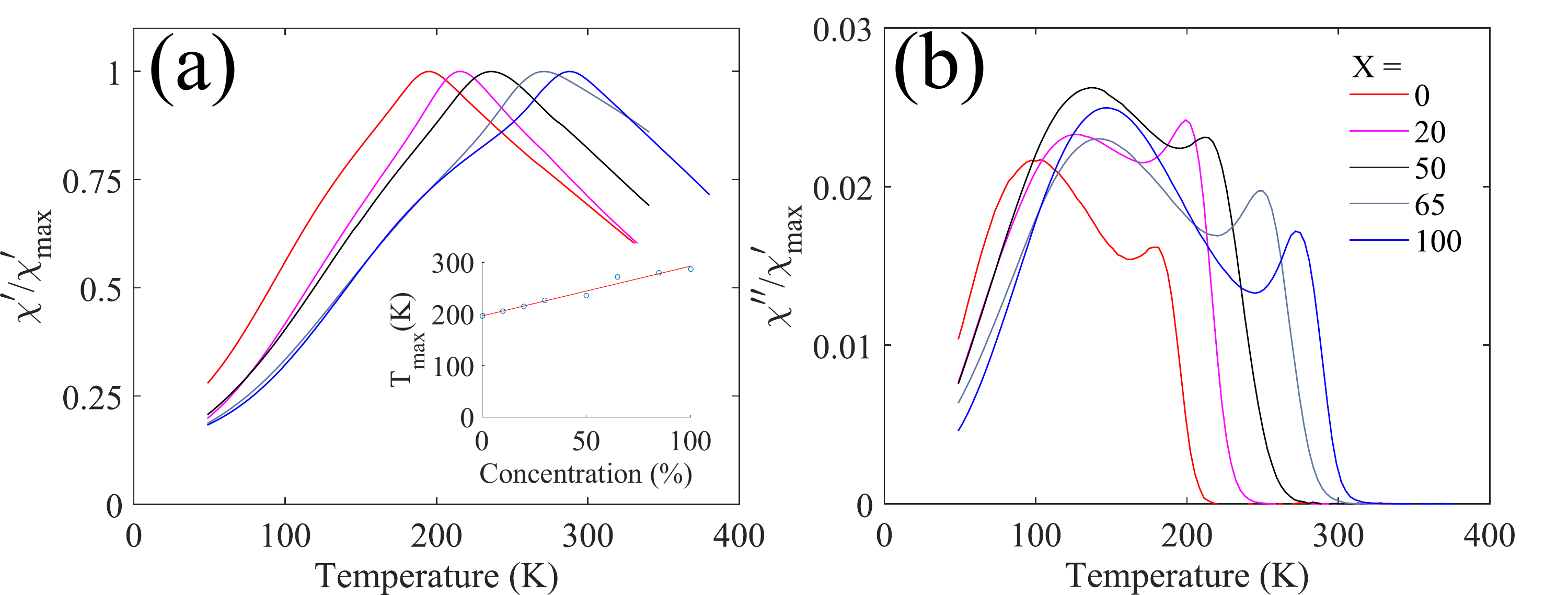}
	\caption{(a) In phase ($\chi^\prime$) and (b) out-of-phase ($\chi^{\prime \prime}$) components of the AC susceptibility as a function of temperature (f=10 Hz) for selected compacts as labeled in the legend in (b). The inset in (a) shows the temperature of the maximum of the in-phase susceptibility ($T_{max}$) as a function of concentration of 11.5 nm particles.} 
	\label{fig:AC_largo}
\end{figure*}

\section{Experiments}
Two highly monodisperse batches of $\gamma$-Fe$_2$O$_3$ nanoparticles of 9 and 11.5 nm diameter were synthesized using the thermal decomposition route described in Ref. \onlinecite{Hyeon2001}. The two batches (still in solution) were mixed into different concentrations, the particles were then collected and the oleic acid removed by repeated washing in acetone. After removal of the oleic acid the particles were dried yielding several powders with different proportions of small/large particles. These powders were pressed into dense compacts (close to random-close-packing\cite{Andersson2015Nanotechnology,Normile2016}) which are referred to as MIXx, where x denotes the percentage by weight of 11.5 nm particles, viz. x=0, 10, 20, 30, 50, 65, 85 or 100. Figs. \ref{fig:TEM_SEM} (a) and (b) show transmission electron microscope (TEM) images of the 9 nm and 11.5 nm  particles, respectively. Their size distributions are shown in panel (d), where it can be seen that there is little overlap between the two distributions. Fig. \ref{fig:TEM_SEM} (c) shows a scanning electron microscope (SEM) image of the MIX65 sample, indicating the random mixing of large and small particles expected from the synthesis method. AC-susceptibility as a function of temperature of all compacts was studied using an AC magnetic field with amplitude $H$ = 80 A/m (1 Oe) and frequency 10 Hz.  The temperature region around the maximum of the in-phase component of the AC-susceptibility curves was studied using several frequencies. The zero field cooled (ZFC) DC magnetization as a function of temperature was measured using an applied field of $H$ = 800 A/m (10 Oe) and a sweeping rate of 2 K/min. DC  memory experiments were performed using a similar protocol to that described in Ref. \onlinecite{Mathieu2001a}, i.e., a 3 h stop was made at 100 K during cooling (zero field) and the subsequent recording of  magnetization (upon heating from base temperature) used $H$ =  800 A/m. All magnetic measurements were performed using an MPMS (5T) SQUID magnetometer from Quantum Design.

\section{Results and discussion}

The temperature dependence of the magnetic AC-susceptibility is shown in Fig. \ref{fig:AC_largo}, where panel (a) shows the in-phase and panel (b) the out-of-phase component. A single, rather sharp, peak is observed in the in-phase component of all compacts. The position of the peak, $T_{max}$, is plotted as a function of concentration of 11.5 nm particles in the inset of Fig. \ref{fig:AC_largo} (a). $T_{max}$ is found to increase linearly with increasing concentration. 
\begin{figure}[t!]
	\centering
		\includegraphics[width=0.4500\textwidth]{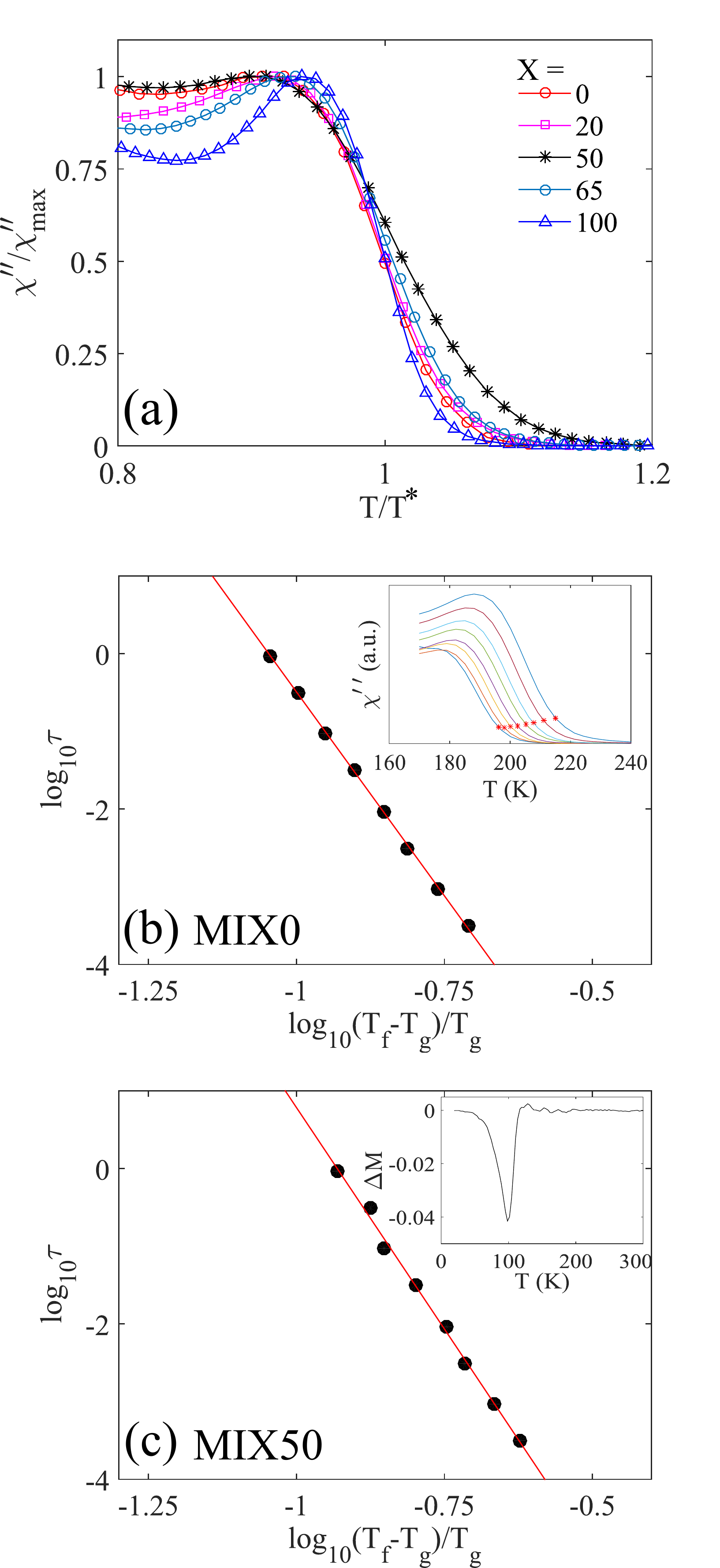}
	\caption{(a) The out-of-phase component component as a function of normalized temperature for all MIXx compacts. The temperature has been normalized to $T^*$, defined as the temperature corresponding to the maximum slope of the onset of dissipation. $\chi^{\prime\prime}$ has been normalized to the maximum close to the onset of dissipation. Fits to a power law, $\tau=\tau_0(T_f/T_g-1)^{-z\nu}$, for (b) MIX0 and (c) MIX50. The inset in (b) shows the out-of-phase component as a function of temperature for several frequencies (0.17 to 510 Hz) for MIX0. The \textcolor{red}{ $\ast$} indicates the determined freezing temperature $T_f$. The inset in (c) shows the result of a memory experiment for MIX50.} 
	\label{fig:AC_sharpness}
\end{figure}

\begin{figure}[t!]
	\centering
		\includegraphics[width=0.45\textwidth]{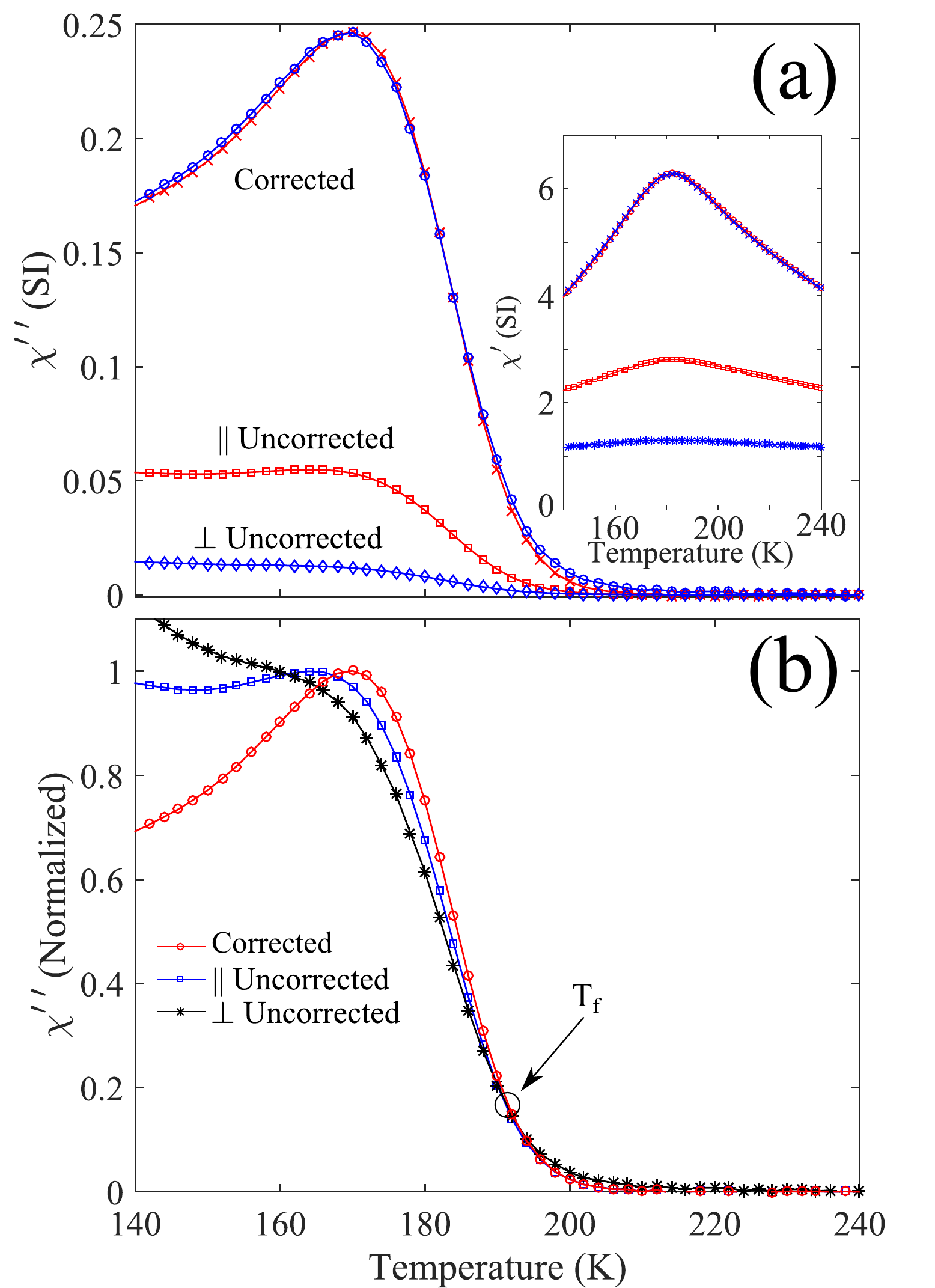}
	\caption{Demagnetization effects in a well defined disc of compacted nanoparticles. (a) Out-of-phase susceptibility as a function of temperature measured with the field perpendicular ($\perp$) and parallel ($\parallel$) to the disc plane as well as demagnetization corrected data for the corresponding curves. Inset in (a) shows the in-phase component as a function of temperature for the same data sets. (b) Normalized out-of-phase component as a function of temperature. The data is normalized by the maximum in $\chi^{\prime \prime}$ at the onset of dissipation. The black circle (1/6 of the maximum for $\chi^{\prime \prime}$) indicates the freezing temperature, $T_f$, used in the critical-slowing-down analyses.} 
	\label{fig:demag}
\end{figure}

As can be seen in Fig. \ref{fig:AC_largo} (b), the out-of-phase component of the AC-susceptibility of all compacts shows a sharp onset near the temperature $T_{max}$. At temperatures below the onset, a wide low temperature maximum appears in the out-of-phase component for all compacts. Similar features were observed in single diameter maghemite NP assemblies of other particle sizes \cite{Andersson2015Nanotechnology} as well as in other NP systems.\cite{Kaman2015,Hansen2002,Parker2008} As shown by Normile et. al,\cite{Normile2016} demagnetization effects can induce broad features in the measured out-of-phase component of the AC-susceptibility, however the temperature of the maximum of the in-phase component is found to be unaffected by such effects. The compacts used in this study did not all possess well defined shapes (e.g., cylindrical or cuboidal), thus a correction of the susceptibility data for demagnetization effects has not been possible across the assembly series. However, the impact of demagnetization effects on the measured susceptibility used for the critical-slowing-down analyses is expected to be negligible, as discussed later (with respect to Fig. \ref{fig:demag}).

In Fig. \ref{fig:AC_sharpness} (a), the out-of-phase susceptibility component normalized to the value of the maximum just below the onset is plotted as a function of reduced temperature, $T/T^*$  (see the caption of Fig. \ref{fig:AC_sharpness}  for a definition of $T^*$). It is seen that the abruptness of the onset curve is least significant in the sample MIX50, and that the degree of sharpness of the absorption onset decreases with increasing mixing (i.e., towards x = 0.5, from higher or lower x). A series of similar dense monodisperse nanoparticle systems of different particle size \cite{Andersson2016MRX} were reported to show similarly sharp onsets of finite out-of-phase components as the current MIX0 and MIX100 compacts. Those particle systems were also found to obey critical-slowing-down as well as exhibiting ageing, memory and rejuvenation phenomena indicating superspin glass behavior (see also Refs. \onlinecite{Hiroi2011APL,Hiroi2011PRB}). Since the onset of the out-of-phase component of MIX50 is less sharp than that of non-mixed systems, it is tempting to ask whether this compact undergoes a SSG phase transition. A DC memory experiment revealed that the sample exhibits ageing, memory [inset Fig. \ref{fig:AC_sharpness} (c)] and rejuvenation, however these properties do not necessarily mean that the system undergoes a phase transition.\cite{Jonsson1995PRL} The AC-susceptibility of all MIXx compacts was measured in an extended frequency window (0.17 to 510 Hz) in narrow temperature intervals near $T_{max}$, see, e.g., the  inset of Fig. \ref{fig:AC_sharpness} (b). Critical-slowing-down analysis\cite{Fischer1991,Hohenberg1977} of the derived freezing temperatures reveals that even the most mixed system (MIX50) exhibits critical slowing down, see Fig. \ref{fig:AC_sharpness} (c). This demonstrates that very strongly interacting magnetic nanoparticle dipolar systems can have broad size distributions and still undergo a SSG phase transition. This result contrasts with relatively "dense" (but much less than the present compacts) ferrofluids, where the comparatively weaker interactions may lead to the collective freezing of only a percolated fraction of the spins in the system.\cite{Parker2008} The critical parameters of all compacts, presented in Table \ref{tab:scaling}, are similar to those reported in Ref. \onlinecite{Andersson2016MRX}  (a study of single diameter NP assemblies). The derived SSG temperatures, T$_g$,  increase linearly with the concentration of 11.5 nm particles, in similarity with the parameter $T_{max}$, see inset of Fig. \ref{fig:AC_largo} (a). Certain ball milled magnetic nanostructured materials with wide size distributions of the magnetic entities exhibit a spin glass transition at low temperatures,\cite{Osth2007,Li1997} however, the interaction mechanism in these systems is direct exchange and/or RKKY and the spin glass behavior is a consequence of competing ferro- and antiferromagnetic interatomic interaction, implying that the relaxation time of the interacting entities remains atomically short down to low temperatures.

\begin{table}[t!]
  \centering
  \caption{Parameters from critical slowing down analyses for all MIXx compacts.}
		\begin{tabular}{l|r|r|l}
  Sample  	&$T_g$ (K) 		&$z\nu$	  & $\tau_0$ (s) 	 	\\\hline
	MIX0 			& 180					& 10 		& 1$\times 10^{-11}$		\\\hline
	MIX10 		& 185					& 10 		& 5$\times 10^{-11}$			\\\hline
	MIX20 		& 200					& 9			& 1$\times 10^{-10}$ 	 	\\\hline
	MIX30 		& 205					& 9 		& 1$\times 10^{-10}$ 		 	\\\hline
	MIX50 		& 215					& 11 		& 1$\times 10^{-11}$  		\\\hline
	MIX65 		& 245					& 10 		& 1$\times 10^{-11}$ 		\\\hline
	MIX85 		& 265					& 10 		& 5$\times 10^{-12}$ 		\\\hline
	MIX100 		& 275					& 9 		& 1$\times 10^{-11}$ 		\\
				\end{tabular}%
  \label{tab:scaling}%
\end{table}%

To investigate a possible influence of demagnetization effects\cite{Normile2016} on the results of the presented critical-slowing-down analysis a random close packed assembly with similar composition (8 nm nanoparticles) and a well defined geometry (disc) was prepared, allowing demagnetization effects on the determination of freezing temperatures to be examined. Data for the out-of-phase component measured with the field in-plane, out-of-plane and data after demagnetization corrections are presented in Fig. \ref{fig:demag} (a). The corresponding in-phase data are plotted in  the inset of Fig. \ref{fig:demag} (a). The data sets show that demagnetization effects are indeed significant. The freezing temperature at a certain frequency is determined as the temperature at which a finite out-of-phase component of the AC-susceptibility (dissipation) first appears. In the current critical slowing down analyses the criterion chosen to define $T_f$ is  the temperature where the out-of-phase component has reached 1/6 of the value at the maximum just below the onset. In Fig. \ref{fig:demag} (b) it can be observed that, the in-plane, out-of-plane and the demagnetization corrected data all indicate the same $T_f$. Therefore, the critical slowing down analysis (Table I) are unaffected by having used uncorrected susceptibility data.

\section{Conclusions}

The magnetic behavior of random dense compacts comprising different ratios of two of magnetic NPs, with the dominant interparticle interaction being dipolar, has been studied. A striking result is that all compacts show a superspin glass phase transition despite the remarkably different particle size distribution (continuously tuned across the assembly series), the size distribution in the most mixed sample (MIX50), in particular, being significantly broader than that of the monodisperse end members of the series (MIX0 and MIX100). The only significant difference in quality of the superspin glass behavior between the monodisperse compacts and MIX50 is a moderate broadening of the width of the onset of the out-of-phase component of the AC-susceptibility. It is found that all compacts exhibit critical slowing down with a spin glass characteristic value of the exponent $z\nu$. The glass transition temperature increases linearly with the concentration of 11.5 nm particles indicating that it is the average interaction strength between all particles that determines $T_g$.

\acknowledgments{
Financial support from the Swedish Research Council (VR),The Research Council of Norway for the support to the Norwegian Micro- and Nano-Fabrication Facility, NorFab, project number 245963/F50, The Institute of Bioengineering and Nanotechnology (Biomedical Research Council, Agency for Science, Technology and Research, Singapore), MINECO (project MAT2015-65295-R) and from the Junta de Comunidades de Castilla-La Mancha (PEII-2014-042-P) is acknowledged.
}

%\end{acknowledgements}

%\bibliographystyle{eplbib}
%\bibliography{mybib}{}
%
\end{document}